 \documentclass[global,twocolumn]{svjour}
\usepackage{graphicx}

\journalname{Applied Physics A}

\begin{document}

\title{Optical absorption and electron energy loss spectra of
carbon and boron nitride nanotubes: a first principles approach.}

\author{A.G. Marinopoulos$^1$, Ludger Wirtz$^2$, Andrea Marini$^2$,
Valerio Olevano$^1$,
Angel Rubio$^2$, and
Lucia Reining$^1$
}
\institute{$^1$Laboratoire des Solides Irradi\'es,
       UMR 7642 CNRS/CEA,
       \'Ecole Polytechnique, F-91128 Palaiseau, France \\
$^2$Department of Material Physics, University of the Basque Country,
Centro Mixto CSIC-UPV/EHU, and Donostia International Physics Center (DIPC),
Paseo Manuel de Lardizabal 4, 20018 San Sebasti\'an, Spain}

\date{Received: date / Revised version: date}

\authorrunning{A.G. Marinopoulos, L. Wirtz, et al}
\titlerunning{Optical properties of C and BN nanotubes}

\maketitle

\begin{abstract}
We present results for the optical absorption spectra of small-diameter
single-wall carbon and boron nitride nanotubes obtained by {\it ab initio}
calculations in the framework of time-dependent density functional theory.
We compare the
results with those obtained for the corresponding layered structures,
i.e. the graphene and hexagonal BN sheets.
In particular, we focus on the role of
depolarization effects, anisotropies and interactions in the excited
states. We show that already the random phase approximation 
reproduces well the main features of the spectra when crystal local field
effects are correctly included, and discuss to which extent the calculations
can be further simplified by extrapolating results obtained for the
layered systems to results expected for the tubes.
The present results are relevant
for the interpretation of data obtained by recent experimental tools for
nanotube characterization such as
optical and fluorescence spectroscopies
as well as polarized resonant Raman scattering spectroscopy.
We also address electron energy loss spectra in the small-q momentum
transfer limit.
In this case, the interlayer and intertube interactions play an enhanced
role with respect to optical spectroscopy. 
\end{abstract}

\noindent Pacs{71.45.Gm,77.22.Ej,78.20.Bh,78.67.Ch}

\section{Introduction}

The science of nanostructures is one of the fields of growing interest
in materials science.
Nanotubes, in particular, are of both fundamental and
technological importance; being quasi-one-dimensional (1D)
structures, they possess a number of exceptional properties.
While the peculiar electronic structure
-- metallic versus semiconducting behavior --
of carbon nanotubes depends sensitively on the diameter and the chirality, 
i.e., on the way the graphene sheet is wrapped up into a tube \cite{dressbook},
boron nitride (BN) tubes display a more uniform behavior with a wide 
band-gap (larger than
4 eV), almost independent of diameter and chirality \cite{bnrubio,bnblase}.
The potential applications of nanotubes in nanodevices are numerous 
\cite{Nano}: super-tough nanotube fibres \cite{Dal}, gas sensors \cite{Modi},
field effect displays \cite{field}, and electromechanical devices \cite{Fenni},
to name only a few.
One of the most spectacular examples is the realization of field effect 
transistors both with carbon nanotubes \cite{Tan,Mar} and with 
BN-nanotubes \cite{Rado}.

Since carbon and BN nanotubes are routinely produced in gram quantities,
the challenge now consists in having fast experimental tools for
characterization of nanotube samples and, if possible, 
single isolated nanotubes. Optical spectroscopic techniques provide us
with this tool~\cite{optical,STS,resraman,bac03,hagen,lefebvre}.
The final goal is to find a unique mapping of the measured
electronic and vibrational properties onto the tube indices $(n,m)$.
To that end, the electronic structure and dielectric properties
of the tubes are two key areas to study.
One possible spectroscopic method is optical absorption
spectroscopy with direct excitations from occupied to unoccupied
states.
For carbon tubes, the energy difference $E_{ii}$ between corresponding 
occupied and unoccupied van Hove singularities (VHSs) in the 1-dimensional 
electronic density of states (DOS) is approximately inversely 
proportional to the tube diameter $d$. In resonant 
Raman spectroscopy~\cite{resraman} and 
scanning tunneling spectroscopy~\cite{STS}, this scaling is employed for the 
determination of tube diameters 
(and $(n,m)$ indexes in resonant Raman~\cite{jorio}).
A recent example is the fluorescence 
spectroscopy of single carbon tubes in aqueous solution,
where $E_{22}$ is probed through the frequency of the excitation laser
and $E_{11}$ is probed simultaneously through the frequency of the emitted
fluorescent light \cite{bac03}. 
Also optical absorption spectroscopy of nanotubes in aqueous solution
\cite{hagen} allows for the spectral resolution of peaks that can
be associated with distinct types of nanotubes.
For the distance between the first VHSs in 
semiconducting carbon tubes, a simple $\pi$-electron
tight-binding fit yields the relation
$E_{11}=E_{22}/2=2a_{C-C} \gamma_0/d$, where $a_{C-C}$ is the 
distance between nearest-neighbor carbon atoms.
The value for the hopping matrix element $\gamma_0$
varies between 2.4 eV and 2.9 eV, depending on the experimental context
in which it is used. This fact is a clear indication that
the above relation gives only qualitative and not quantitative information
on the tube diameter and/or chirality. Furthermore, for small-diameter
tubes the band structure completely changes with respect to the band
structure of large diameter tubes, including a reordering of the VHSs in
the density of states and displaying fine structure beyond the first 
and second VHSs~\cite{tube_prl,kirchberg}. 
This structure sensitively depends on the tube 
indices and may be probed by optical absorption spectroscopy over a wider 
energy range, possibly extending into the UV regime.

The spectroscopical characterization of macroscopic tube samples
is made difficult by the fact that in the bulk solid and even in bundles the
tubes are not perfectly aligned and do not have a well defined
diameter and helicity, rendering, in the case of carbon tubes, a random mixture 
of semiconducting and metallic tubes \cite{exp}.
Additionally, the role of intertube interaction in the spectra
needs to be addressed since the
tubes are close packed and could interact with each other via
long-range forces induced by the excitations.
Only very recently optical absorption
spectra were reported for aligned single-wall carbon tubes of a very 
narrow diameter 
distribution (4$\pm$0.2~\AA) grown inside the channels of a zeolite matrix 
\cite{Li01}. Geometric arguments predict three
possible tubular helicities for the range of diameters around 4 \AA:
the armchair (3,3), the zig-zag (5,0) and the chiral (4,2) configuration.
Therefore, this particular case serves as an important case study where
a direct comparison between experimental data and theoretical
calculations becomes possible. Indeed in Ref.~\cite{tube_prl}
we have shown the relevance of a first-principles calculation
for these small-diameter carbon nanotubes by reproducing
the polarization dependence of the measured optical spectra.

At present, the dielectric response of tubes in the frequency range
of the electronic interband transitions and the 
collective excitations (plasmons)
of the valence electrons (up to 50 eV) is not well understood since
previous studies focused on the low-energy regime 
and excitations~\cite{And,lowE}. For higher frequencies,
up to now, the majority of the theoretical studies of the response,
besides model calculations \cite{mode},
are mostly limited to summing over independent band-to-band transitions
obtained within the semi-empirical tight-binding method \cite{Lin} or
the density-functional theory (DFT) framework \cite{Li01,DFT1,DFT2},
or to calculations of the
joint density of states (JDOS) \cite{jdos} which use the
bandstructure with no explicit evaluation of the transition matrix elements.
As we will show below, these approximations are
not sufficient for a full interpretation of the experiments.
This shortcoming is not due to the quality of the bandstructure calculation
itself but instead due to the neglect of
the induced microscopic components in the response to the
external field, the local field effects (LFE)~\cite{tube_prl}. These effects
strongly modify the total response for certain polarizations of the 
external perturbation. Also, induced exchange and correlation
(XC) components obtained beyond the random-phase approximation (RPA)
might contribute \cite{And,REV}.
Therefore, important questions concerning the
electron interaction, excitations and screening still remain unanswered.
Our approach here is to determine
the spectra by {\it ab-initio} calculations incorporating important
ingredients of the electron response as in previous
works~\cite{tube_prl,kirchberg}.

The present paper is organized as follows:
after an exposition of the theoretical framework in section \ref{theory},
we present in section \ref{tubeabs} {\it ab initio} calculations of optical
spectra of small single-wall carbon and boron nitride nanotubes. We identify
the influence of crystal LFE, XC effects and intertube interaction in the
spectra.
We also point out the similarities and differences between carbon and BN
structures which are related to similarities and differences
in the respective electronic band structures.
A certain number of comparisons with experiment allows us to
verify that the chosen {\it ab initio} approach is an important 
improvement and well suited for
the description of the spectra of these systems.
In section \ref{sheetabs}, we analyse the spectra of the building
blocks of the tubes, i.e. the graphene and hexagonal BN sheets (layers)
including the optical spectrum of graphite.
This gives information about the
the interaction between objects (sheet-sheet)
in the excited state, even when these
objects can be considered to be isolated in the ground state.
We also discuss to which extent the dielectric response of the tubes
can be understood in terms of the response of the sheets.
This will help to understand whether some aspects of the response of the
tubes are {\it inherent} to the sheets, and if so, they could also
be observed in other systems of more practical interest, e.g.
samples comprising a mixture of tubes with different diameters
and orientation or multiwall tubes.
In section \ref{eelsec}, we present results for electron energy
loss in order to have some additional validation from existing
experimental data, and because a comparison between optical and loss spectra
allows a deeper discussion and understanding of interaction effects.
Finally, we conclude with section \ref{finsec}, with an overall discussion
of the results.

\section{Theoretical Framework}
\label{theory}

{\it Ab initio} calculations in the framework of Density Functional Theory 
(DFT) have yielded high-quality results for a large variety of systems:
from molecules to periodic solids and structural defects~\cite{dft_book}.
These results are however mostly limited to quantities related to
the electronic {\it ground state}, whereas additional phenomena that 
occur in the excited state are not correctly described~\cite{self}.
In particular, the self-consistency between the total
perturbing potential and the charge response induces Hartree and
exchange-correlation potentials that have to be dealt with. The former
give rise to the so-called local field effects, whereas the latter
can lead, for example, to excitonic effects.
Today, in the solid state {\it ab initio} framework
two main approaches can include both LFE and XC
effects~\cite{REV} and can be therefore suitable for the description of
electronic excitations in nanostructures.
First, Green's functions approaches within many-body perturbation theory:
here one adds self-energy corrections
to the DFT Kohn-Sham bandstructure and the electron-hole interaction
is included via the solution of the Bethe-Salpeter equation \cite{REV}.
This approach has given excellent results for various bulk and finite
systems, but it is computationally very cumbersome
and not ready yet to be applied systematically to more
complex systems.
The time-dependent Hartree-Fock (TDHF) method represents a certain level
of approximation within this approach, where correlation (i.e. screening)
in the self energy is neglected. LFE, on the other hand, are still
retained. TDHF has been employed, up to now in a semi-empirical way,
to calculate optical spectra, including excitons,
of carbon nanotubes~\cite{And} and also
for the 4 \AA-diameter ones \cite{TDHF}.
However, the obtained absorption peak assignments in the latter case
were in disagreement with predictions based on the operative dipole
selection rules for the specific tubular space groups \cite{Li01,DFT1}.
This suggests that in such systems with important metallic character
a neglect of screening may lead to problems.

The second way to calculate spectra
is given by the time-dependent DFT (TDDFT)~\cite{RGK}, where all manybody
effects are embodied in the exchange-correlation potential and kernel.
It is in practice an approximate way
but always treats the variations of the Hartree potential exactly (right
as in both the Green's functions and TDHF approaches).
This approach, using the (static) adiabatic local
density approximation (TDLDA) for the XC effects, or even completely
neglecting the density variations of the XC potential (RPA, on the
basis of an LDA bandstructure) has been applied
successfully to many finite and infinite systems~\cite{REV}. In particular,
excellent results for the loss spectra of graphite have been obtained
using this approximation~\cite{gra1}.
This is the approach we follow in the present work.

As a first step in our TDDFT approach, we determined the electronic
ground state of the systems. The Kohn-Sham single-particle equations were
solved self-consistently in the LDA for exchange and correlation~\cite{HOKS}.
For the description of the valence-core interaction we have used
norm-conserving pseudopotentials which were
generated from free-atom all-electron calculations \cite{TM}.
For the ``isolated'' geometries of the tubes and of the sheets we
calculated periodic arrays with a large distance between the building
blocks in a supercell geometry,
in order to minimize the mutual interaction.
The crystalline valence-electron wavefunctions were expanded using
a plane-wave basis set (with an energy cutoff of 62 Rydberg for carbon
systems, and 50 Rydberg for BN systems). A part of the calculations 
was carried out using the ABINIT code \cite{abinit}.

The next step is the linear response calculation of the independent particle 
polarizability $\chi^0$ \cite{EHR59,notechi}. It involves a sum over 
excitations from occupied bands to unoccupied bands:
\begin{eqnarray}
&\chi^0_{{\bf G},{\bf G'}}({\bf q},\omega) = 
2 \int \frac{d k^3}{(2\pi)^3} \sum_n^{occ.} \sum_m^{unocc.}  \nonumber \\
&\left[\frac{\langle n,{\bf k}|e^{-i({\bf q}+{\bf G}){\bf r}}|m,
                                                   {\bf k}+{\bf q}\rangle
     \langle m,{\bf k}+{\bf q}|e^{i({\bf q}+{\bf G'}){\bf r}}|n,{\bf k}\rangle}
     {\epsilon_{n,{\bf k}} - \epsilon_{m,{\bf k}+{\bf q}} - \omega \ - i\eta}
    - (m \leftrightarrow n) \right],
\label{chi0}
\end{eqnarray}
where $(m \leftrightarrow n)$ means that the indices $m$ and $n$ of the
first term are exchanged.
The result is checked for convergence with respect to the number of bands
\cite{nvalues} and the discrete sampling of {\bf k}-points within the first
Brillouin zone.

Within TDLDA, the full polarizability $\chi$ is connected to $\chi^{0}$ 
via\cite{Pet96}
\begin{eqnarray}
\label{eq:chi}
      \chi = \chi^{0} + \chi^{0} (V_{C}+f_{xc}) \chi ,
      \end{eqnarray}
where V$_{C}$ is the bare Coulomb interaction
and $f_{xc}$, the so-called exchange-correlation kernel, is the
functional derivative of the LDA exchange-correlation potential with
respect to the electron density.
By setting $f_{xc}$ to zero, exchange and correlation effects in the
electron response
are neglected and one obtains the Random-Phase Approximation (RPA).
We have carried out our calculations in the RPA and also in the TDLDA
for certain cases.
The inverse dielectric function for a periodic system and momentum
transfer ${\bf q}$ is obtained from:
\begin{eqnarray}
      \varepsilon^{-1}_{{\bf G},{\bf G}'}({\bf q}) = \delta_{{\bf G},{\bf G}'}
       +V_{C}({\bf q}+{\bf G}){\chi}_{{\bf G},{\bf G}'}({\bf q})
      \end{eqnarray}
with ${\bf q}$ in the first Brillouin zone and
${\bf G}$, ${\bf G}'$ are reciprocal lattice vectors.
The absorption spectrum is obtained as the imaginary part of the macroscopic
dielectric function
\begin{eqnarray}
       \label{LFE}
       \varepsilon_{M}(\omega ) = 1/\varepsilon_{{\bf G}={\bf 
G}'=0}^{-1}({\bf q}
       \to 0;\omega )
\end{eqnarray}
whereas the loss function
for a transferred momentum $({\bf q}+{\bf G})$
is given by $-{\rm Im}[\varepsilon^{-1}_{{\bf G},{\bf G}}(\bf q)]$.
In inhomogeneous systems, e.g. periodic solids, clusters and structural
imperfections the inhomogeneity in the electron response gives rise to
local field effects (LFE) and the $\varepsilon_{{\bf G},{\bf G}'}$ cannot be
considered as being purely a diagonal matrix. Therefore,
its off-diagonal elements have to be included in the matrix inversion.
Making the approximation $\varepsilon^{-1}_{{\bf G},{\bf G}}(\bf q)
\approx 1/\varepsilon_{{\bf G},{\bf G}}(\bf q)$ corresponds to
neglecting the inhomogeneity of the response, i.e. to neglecting the LFE.
When these effects are neglected and, moreover,
all transition matrix elements in $\chi^{0}$ are supposed to be constant, one
arrives at the widely used approximation that the absorption spectrum
is proportional to the JDOS, i.e. proportional to the sum over
interband transitions from
occupied ($v$) to empty ($c$) states over the Brillouin zone
points ${\bf k}$,
$\sum_{v,c,{\bf k}} \delta (\epsilon_{c{\bf k}}-\epsilon_{v{\bf
k}}-\omega )$.

\section{Optical Absorption Spectrum for carbon and BN tubes}
\label{tubeabs}

Let us first look at the optical absorption spectra of 
small-diameter carbon nanotubes
(all three possible helicities), for
which experimental results have recently become available.
The calculations of the spectra \cite{calcc} 
were done for the tubes arranged in a
hexagonal lattice with an intertube distance
(distance between tube walls), equal to $D_{t}=$ 5.5 \AA~
which leads to nearly isolated tubes. Additionally,
for the (3,3) armchair ones we repeated the calculations
for a solid with
a smaller intertube distance, $D_{t}=$ 3.2 \AA, which is close
to the interlayer distance in graphite ($\sim$ 3.4 \AA).

\begin{figure}[h]
\includegraphics*[width=8cm]{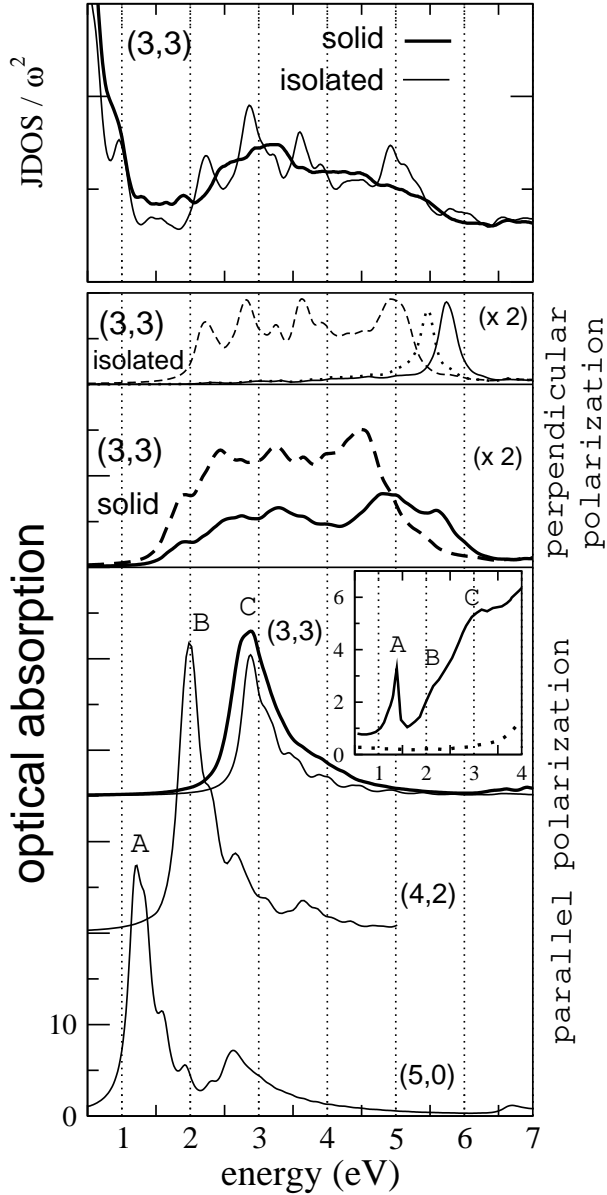}
\caption []
{Calculated optical spectra for the (3,3), (5,0) and (4,2)
nearly isolated and solid (only for (3,3) case; thick lines) of carbon tubes 
with (continuous
lines) and without (dashed lines) LFE in the RPA. For comparison
the TDLDA result for the isolated (3,3) tubes is given by the dotted line
in the second panel as well as the JDOS curves.
The results
are presented for both perpendicular and parallel to the tube-axis polarization.
For the (4,2) tube only the result for the parallel
polarization without LFE is given, which should be a sufficiently good 
approximation for this polarization, according to our results for 
the (3,3) and (5,0).
The experimental data \cite{Li01} are displayed in the inset.}
\label{fig1}
\end{figure}

The optical absorption spectra for the small diameter 
tubes are diplayed in Fig.~\ref{fig1}
(as well as JDOS curves for two cases).
In the upper panel of the Figure, the JDOS divided by the
square of the excitation energy, for the (3,3) nearly isolated
(thin line) and interacting (in the solid) (thick line) tubes are
shown.  It can be seen that after an initial very steep decrease
($<$~1 eV),
in both JDOS curves there is a gradual increase starting from 2 eV.
In the case of isolated tubes a sequence of
pronounced peak structures up to 5 eV is observed: this can be explained
from the occurrence of direct interband transitions between the
van Hove singularities
of the density of states (DOS). These peaks are smeared out
in the JDOS of the solid where the tubes are strongly
interacting \cite{ANGEL}.
In the lower panels of Fig.~\ref{fig1} the calculated absorption spectra
for light polarizations
perpendicular and parallel to the tube axis are displayed (as well as
the experimental data plotted in the inset). The dashed
lines denote results in the RPA neglecting LFE,
continuous lines including LFE.
For parallel polarization both LFE and LDA-XC effects were found to be
negligible and, therefore, in this case only the RPA results without
LFE are presented.
The fact that both these effects turned out to be negligible can explain
why the peak positions (A, B and C) predicted here for the parallel
polarization match very well the ones found in two recent DFT-RPA
studies \cite{DFT1}, which completely neglected LFE and XC effects in the
response.
Our calculated peak positions (A, B and C) in Fig.~\ref{fig1} are
in good agreement (to within 0.2 eV) with the
experimentally observed peak structures for the parallel
polarization (inset).

It is also important to note that the present results support the same
peak-to-helicity assignment as in previous works \cite{Li01,DFT1}.
Namely, the three observed peaks -- A, B and C -- are due to the (5,0),
(4,2) and (3,3) tubes, respectively. Such an assignment is also in
accordance with the dipole selection rules for these tubular space groups 
\cite{Li01,DFT1}.

Concerning the effect of intertube interaction in the absorption
we found that it is rather small for this polarization:
tube-tube interaction leads only to a broadening of the main absorption
peak (thick solid curve for the (3,3) tubes).
This can be explained from an increase of
the energy range of the possible interband transitions
brought about by the interaction.

All the discussion of the results up to now seems to indicate that LFE
may not be needed. Nonetheless, for light polarizations perpendicular
to the tube axis -- due to the presence of depolarization -- LFE play
an important role which cannot be ignored. More specifically, for this
polarization the experimental spectrum displays vanishing intensity for
frequencies up to 4 eV \cite{Li01} (see inset in Fig.~\ref{fig1}, dotted curve).
Clearly, neither the JDOS calculation nor the calculated spectrum
without LFE can capture this effect. Instead, they both predict a series
of peaks from 2 to 5 eV. The reason that RPA-without-LFE fails here
is due to the depolarization effect \cite{tube_prl,And,BENE}
 which is created by the
induced polarization
charges. The depolarization is only accounted for if LFE are included:
as it can be seen in Fig.~\ref{fig1} (second panel; continuous curve)
{\it LFE suppress the low-energy
absorption peaks} and render the
tubes almost transparent below 5 eV in agreement with the experiment.
It should not come, therefore, as a surprise that the recent DFT-RPA
studies \cite{DFT1} (which did not consider LFE) did not reproduce
this transparency for the perpendicular polarization.

The TDLDA result, displayed as the dotted curve in the second panel,
turned out to be qualitatively similar to the RPA-with-LFE result:
again the low-energy absorption peaks are suppressed. This shows that
the main effect comes
from fluctuations of the Hartree-, and not from those of the XC-potential
(as also
obtained for the case of graphite~\cite{gra1}).
LDA-XC effects only cause a small (0.3 eV) redshift of the remaining absorption
peak to 5.5 eV,
with respect to RPA-with-LFE. This is a characteristic behavior of finite
systems \cite{REV}. Still there is an open
question about the role of electron-hole interaction in the case of the
small band-gap (4,2) tube \cite{excilou}, however the good agreement with
experiment indicates that its contribution should be small for the 
parallel polarization.

LFE also have a similar drastic impact for the system of isolated
zig-zag (5,0) tubes (not shown).
The manner according to which LFE operate may be understood as follows:
for perpendicular polarization
the tubes form an assembly of almost
isolated, but highly polarizable, objects. An applied external
field induces hence a local, i.e. microscopic, response - the LFE -
which strongly weakens the total perturbation (i.e. it is a
``depolarization''). The
macroscopic response to this weak perturbation is only very moderate,
because the electrons are localized on the tubes. This is totally different
from the screening in
a bulk metal or small-gap semiconductor, where even a very
weak total perturbation still
leads to a strong response at low frequencies. For polarization parallel to
the tube axis, the situation resembles rather this latter case, something
which explains the absence of LFE for this polarization.
On the other hand -- for the perpendicular polarization -- when the tubes
are interacting in the solid (third panel) the depolarization is much
weaker: the tubes start to absorb (they are no longer transparent)
because the electronic states start to delocalize and the system is now
more similar to a bulk metal.
This different behavior of the response depending on the intertube
distance leads to a very important consequence:
it suggests that {\it the inter-tube
interaction can be detected experimentally} in a qualitative study of
the absorption
spectra for perpendicular light polarizations.

\begin{figure}[h]
\includegraphics*[width=8cm]{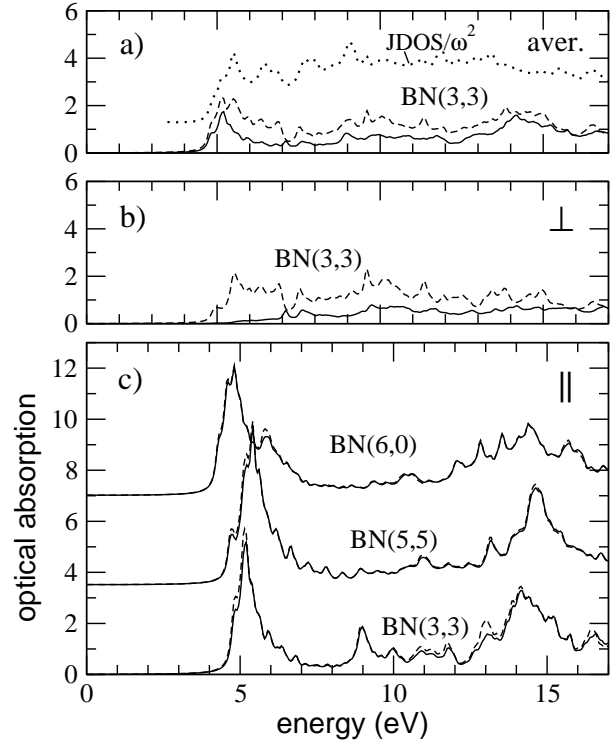}
\caption []
{Calculated optical absorption spectra for quasi-isolated BN nanotubes 
\cite{calcbn} with (solid line) and without (dashed line) LFE in the RPA 
approximation:
a) Comparison of the spatially averaged spectrum of the (3,3) tube
with the joint density of states divided by the square of the
excitation energy (dotted line); 
b) spectrum with polarization perpendicular to the tube axis for the (3,3) tube;
c) spectra with polarization parallel to the tube axis for the (3,3), (5,5)
and (6,0) tubes.}
\label{bntubesfig}
\end{figure}
In Fig.~\ref{bntubesfig} we present optical absorption spectra for
three different BN nanotubes.
First, we compare for the case of the BN(3,3) tube the spatially averaged
spectrum with the joint density of states (JDOS), divided
by the square of the transition energy. If local field efffects are neglected, 
most peaks of the JDOS are visible in the averaged absorption spectra while 
some peaks are suppressed due to small or vanishing oscillator strength in 
Eq.~(\ref{chi0}). This demonstrates that the fine structure in the
spectra is not an artifact of low k-point sampling but is due to the 
presence of van-Hove singularities in the 1-dimensional density of states of 
the tubes. Proper inclusion of LFE leads to a smoothing of the spectrum. 
However, some fine structure survives and may be discernible in high-resolution
optical absorption experiments.
The spectrum with perpendicular polarization demonstrates that for
BN nanotubes, depolarization effects play a similarly important role
as for carbon tubes: Neglecting LFE, the onset of absorption would be
around 4.5 eV.
LFE, however, lead to a redistribution of the
oscillator strength to higher energies and render
the tubes almost transparent up to 8 eV. 
The band gap of BN nanotubes in DFT-LDA is 4 eV and is only weakly dependent 
on radius and chirality \cite{bnrubio,bnblase}.
Accordingly, the absorption spectra for polarization parallel to the
tube axis are very similar for the (3,3), (5,5), and (6,0) tubes
which all display a strong absorption peak around 5 eV and a second
high peak around 14 eV. We expect these structures to be stable also
for tubes with larger diameters. Only the fine structure of the spectra 
(e.g., the absorption peak at 9 eV for the (3,3) tube) depends on the
details of the band-structure and varies for the different diameters
and chiralities~\cite{kirchberg}.
For the (6,0) tube, the first high absorption peak
is split and the dominant peak shifted towards lower energy.
This is a curvature effect which leads to a reduction of the band-gap
for small zigzag BN tubes.

In the next section we will see to which extent the above findings 
for carbon and BN nanotubes can be understood by an analysis based on results 
for graphene and BN sheets and graphite.

\section{Optical Absorption Spectrum for graphite and
the graphene and BN sheets}
\label{sheetabs}

Hexagonal graphite has the ABA Bernal stacking sequence of the graphene
sheets. In the present work we assumed the experimental lattice parameter
a$_{hex}$ and (c/a)$_{hex}$ ratio (2.46 \AA~ and 2.73, 
respectively \cite{latt}).

\begin{figure}[h]
\includegraphics*[width=8cm]{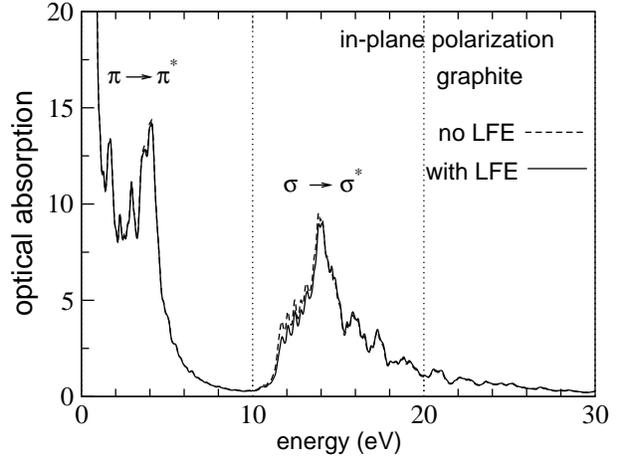}
\caption []
{Absorption spectrum of graphite for {\bf E} $\perp$ {\bf c}. The star
symbol denotes the unoccupied electron states.}
\label{graphabs}
\end{figure}

\begin{figure}[h]
\includegraphics*[width=8cm]{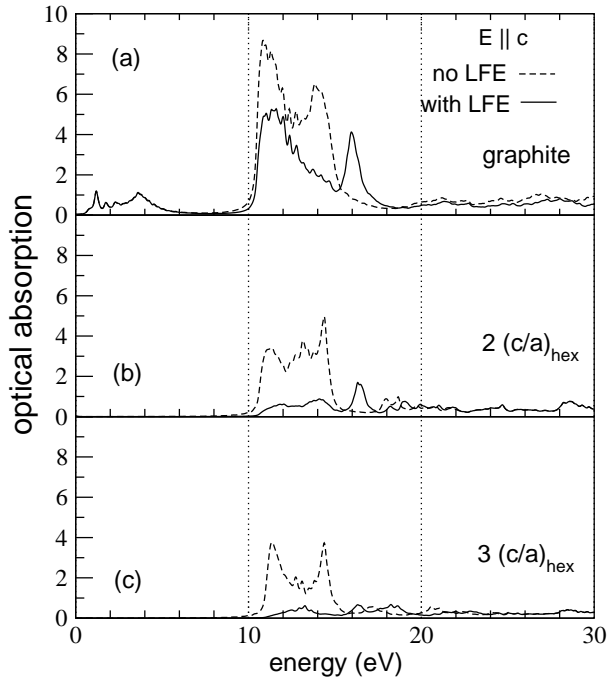}
\caption []
{Absorption spectrum for {\bf E} $\parallel$ {\bf c}, for graphite and
the graphene-sheet geometries with 2 (c/a)$_{hex}$ and 3 (c/a)$_{hex}$.}
\label{fig4}
\end{figure}

The calculated RPA optical absorption spectra \cite{footcalc}, 
with and without LFE,
are shown in Fig.~\ref{graphabs} for in-plane light polarization
({\bf E} $\perp$ {\bf c}) and in Fig.~\ref{fig4} (a) for polarization
parallel to the {\bf c} axis ({\bf E} $\parallel$ {\bf c}).
The in-plane spectrum is dominated by a very intense peak structure at
low frequencies (up to 5 eV) and also another peak structure of
broader frequency range which sets in beyond 10 eV and
has a pronounced peak at 14 eV.
The origin of these peak structures is due to $\pi \to \pi^{\star}$ and
$\sigma \to \sigma^{\star}$ interband transitions, respectively, according
to the earlier interpretations by Bassani and Paravicini \cite{BAS67}
who assumed
a two-dimensional approximation -- no interaction between the graphene
sheets --  and the operative dipole selection
rules for this polarization.
Our calculations of the oscillator strength
for specific transitions
between bands in the Brillouin zone (BZ) are in agreement with their
interpretation.

LFE are found to be nearly negligible for this polarization. This is not
surprising since for in-plane polarizations graphite is homogeneous
in the long-wavelength limit ($q \to 0$).
The general aspects of the spectrum -- peak positions, their intensity and
lineshape -- are in
close agreement with the existing experimental results \cite{exp2} and
the previous
all-electron calculation of Ahuja et al. \cite{AHU97} who neglected LFE.

The absorption spectrum of graphite for the light polarization parallel to
the {\bf c} axis is shown in Fig.~\ref{fig4} (a). 
It is characterized by a weak intensity
in the low frequency range (0-5 eV) and important peaks in the
frequency range beyond 10 eV.
For this polarization the bandstructure does not play the
exclusive direct role in defining the absorption spectrum.
Now LFE are very important.
When LFE are not considered the peak positions for this polarization
are at 11 and 14 eV as in the earlier DFT-RPA calculation \cite{AHU97}.
However, when LFE are included transitions are mixed and
the absorption spectrum is appreciably
modified. The main effect of local fields is
to shift the oscillator strength at 10-15 eV to higher frequencies.
They decrease the intensity of the 11 eV peak and are responsible for
the appearance of the 16 eV peak in the spectrum.
The latter peak is seen in experiments as a shoulder
suggesting that the
inclusion of LFE is necessary.

The non-zero
oscillator strength found below 5 eV is attributed to the inter-layer
interaction which is present in the solid. It is also observed experimentally.
The dipole selection rules \cite{BAS67} for
an isolated graphene sheet (i.e. complete two-dimensionality) lead
to vanishing matrix elements and oscillator strength at this frequency range.
LFE do not have any influence on
the lower part of the spectrum (less than 10 eV).

The existing experimental evidence is not conclusive
for the dielectric function in the 11 eV frequency region.
The frequency-dependent Im[$\varepsilon_{M}$]
obtained from electron energy loss data \cite{Zep-th,ToBa,Veng}
displays a very sharp and intense peak (Im[$\varepsilon_{M}$]=10) at 11 eV.
On the other hand, on the basis of optical measurements \cite{KLU74}
the observed maximum at this frequency is of considerably smaller intensity.
The earlier interpretations \cite{band}
were based on semi-empirical tight-binding bandstructure
calculations in the two-dimensional approximation
(i.e. isolated graphene layers) and they predicted peak structure
between 13.5 and 16.5 eV.

It is therefore also important to understand the effect of the inter-layer
interaction in the optical response and if this interaction is primarily
responsible for the occurrence of the intense peak at 11 eV.
For this purpose, we progressively increased the inter-layer spacing
(doublying and tripling the (c/a)$_{hex}$ ratio). This yields stackings
of graphene layers in the unit cell with much weaker mutual interaction.
The absorption spectra for these graphene-sheet geometries are displayed
in Fig.~\ref{fig4} (b,c).
The first observation is that
the oscillator strength vanishes completely in the frequency region
below 10 eV in complete accordance with the predictions based on the
dipole-selection rules for an isolated graphene layer.
Therefore, the double inter-layer spacing leads to
non-interacting graphene layers as far as the RPA
absorption spectrum (at this frequency range) is concerned.
When LFE are neglected,
the peak structure in the 10-15 eV range stays intact, now with a
smaller intensity due to the larger volume.
With increasing interlayer separation, LFE become progressively more
important. The shift of oscillator strength induced by LFE is so big
that the absorption peaks at 10-15 eV are almost completely suppressed.
These findings demonstrate that both inter-layer interaction and
LFE influence considerably the intensity of the absorption
peak at 11 eV.
Qualitatively, the depolarization effects found in the tubes can hence be
explained by the local field effects observed in the
graphitic response perpendicular to the graphene sheets. Likewise,
the effect of intertube interaction in the spectra for polarizations
perpendicular to the tube axis is also consistent with the 
increased propensity of LFE to shift oscillator strength at higher
frequencies when the intersheet distance progressively increases 
(see Fig.~\ref{fig4}).

\begin{figure*}[ht]
\includegraphics*[width=1\linewidth]{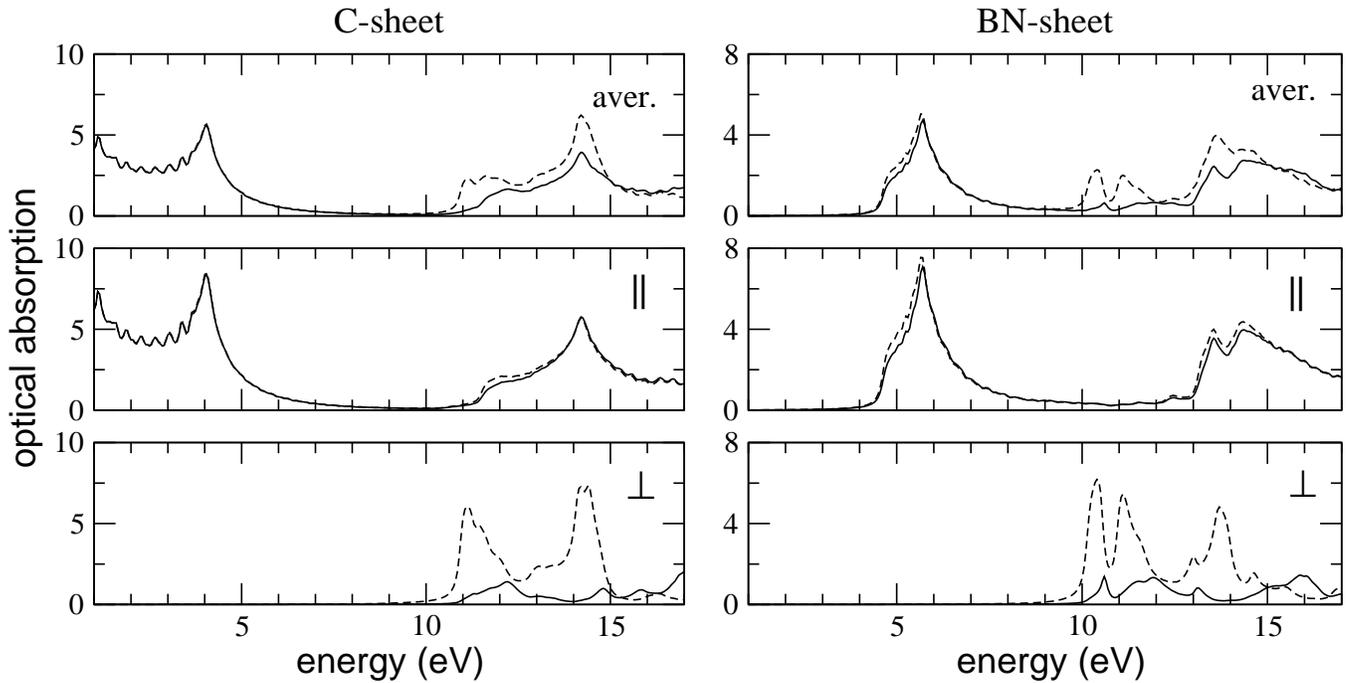}
\caption []
{Calculated absorption spectra for the graphene and hexagonal BN 
sheets~\cite{sheetcalc}.}
\label{sheetsfig}
\end{figure*}
We conclude this section with a comparative 
presentation of the absorption spectra
of the graphene and BN sheets using the same inter-sheet distance 
for both cases, shown in Fig.~\ref{sheetsfig}.
Also we make comparisons with the spectra of the corresponding tubes.
The spectrum of the graphene sheet for the in-plane polarization 
resembles closely the in-plane polarization spectrum of graphite 
(Fig.~\ref{graphabs})
(except for a scaling due to the change in the volume),
confirming once more, as in the case of the tubes for the polarization
parallel to the tube axis,
that the position of the absorption peaks and their lineshape
is only weakly influenced by the distance and inter-sheet
interaction in this case.
The main difference between the graphene and BN-sheet spectra for the in-plane
polarization is the complete absence of any feature below 4 eV in the BN 
spectrum. This is clearly related to the 4 eV LDA-band gap of BN.
The out-of-plane polarization spectra of the graphene and BN sheets
are remarkably similar. Most importantly, both display a transparency
up to 10 eV and both demonstrate a strong depolarization effect with
shift of oscillator strength to higher energies \cite{kirchberg}.

We compare now the calculated sheet-absorption spectra
for in-plane polarization with the corresponding 
tube-absorption spectra for polarization parallel to
the tube axis (Fig.~\ref{fig1} for carbon and Fig.~\ref{bntubesfig} for BN).
In the carbon case, the strong absorption feature
of the sheet between 0 and 4 eV maps onto one or several
absorption peaks of the tubes in this energy range. The details of this
mapping are, however, sensitively dependent on the diameter and the chirality 
of the tubes, since dipole selection rules play a strong role in these 
highly symmetric systems. In particular, for tubes with larger diameter
than the ones calculated in this article, the distance $E_{11}$ of the
first van Hove singularities strongly depends on whether the tube is metallic
or semiconducting.
Therefore, calculations on graphite or graphene alone will not be sufficient
to predict absorption spectra of small-diameter carbon nanotubes.
For BN, in contrast, the first high absorption peak at 5.5 eV maps
directly onto a corresponding peak in the tubes. Also the high energy
absorption feature around 14 eV is very similar for the BN sheet
and the BN tubes. Due to the large band gap of BN, the comparison
between sheet and tube spectra is much more favourable for BN than
for carbon tubes.

On the other hand, in the case of carbon, the extrapolation of results 
from graphite and graphene to the
tubes is much more straightforward for the electron energy loss spectra 
as discussed in the following section.

\section{EELS spectra in the limit of small q: tubes versus the layered 
structures}
\label{eelsec}

Having seen the strong interaction effects that occur in optical spectra
for a polarization perpendicular to the planes or tube axis, it is
interesting to make an excursion to a different type of spectroscopy,
namely electron energy loss (EELS) \cite{fink}. 
Although also in this case one can measure,
like in the absorption experiment, a
momentum transfer close to zero, there is a substantial difference in the
definition of the response function: in the absorption measurement, one
detects the response to the total macroscopic field, whereas in EELS the
response to the external field is reported. Therefore, as pointed out
above, absorption is linked to the imaginary part of the macroscopic
dielectric function, but EELS is linked to the imaginary part of the
inverse of the latter. Mathematically this translates into the fact that
the crucial response function for EELS is governed by equation
(\ref{eq:chi}), whereas in the case of absorption one can use slightly
different
quantity: the macroscopic dielectric function can in fact be rewritten as
\begin{equation}
\label{eq:epsmac}
\varepsilon_M (\omega) = 1 - lim_{{\bf q}\to {\bf 0}} V_C({\bf q},G=0)
\bar \chi_{G=G'=0} ({\bf q};\omega)
\end{equation}
where V$_C(G=0)$ is the long-range component of the Coulomb potential, and it
has a $1/q^2$ divergence for vanishing $q$ and
the quantity $\bar \chi$ obeys an equation similar to (\ref{eq:chi}): 
$\bar \chi = \chi^{0} + \chi^{0} (\bar V_{C}+f_{xc}) \bar\chi$,
but setting the $G=0$ component of $V_C$ to zero ($\bar V_{C}$)~\cite{REV}.
Therefore, this seemingly tiny
difference is responsible for the difference between, e.g., the position of
the main absorption peaks and that of the valence plasmons in solids, and
one can also expect that it will be crucial when interaction effects on the
spectra are discussed. In particular, EELS spectra should show, due to the
presence of this long-range term, stronger interaction effects than
absorption spectra.

\begin{figure}[h]
\includegraphics*[width=8cm]{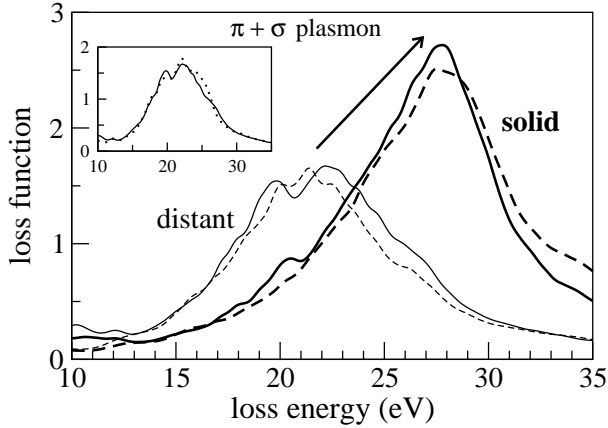}
\caption []
{Calculated RPA loss function for the carbon (3,3)
distant tubes (continuous thin line) and solid (continuous thick line) of tubes.
{\bf q} is vanishing with an orientation parallel to the tube axis.
The dashed curves denote the results for graphite (thick) and graphene (thin)
for small in-layer {\bf q} \cite{gra1}. The loss function for the (5,0) tube
is shown in the inset as a dotted curve. LFE are negligible for this
{\bf q} orientation.}
\label{ctubeels}
\end{figure}

This is in fact the case, as we will illustrate in the following
for the total $\pi + \sigma$ plasmon in graphite. This plasmon
represents the collective excitation mode of all the valence electrons
in graphite.
Before discussing the results we stress here that
the tubes cannot be considered as completely isolated objects in this
calculation of the
loss function with the 5.5 \AA~intertube distance. We refer to the tubes in
the latter geometry as {\it distant}.
Fig.~\ref{ctubeels} shows the RPA loss function,
$-Im(1/\varepsilon_{M})$, for
the (3,3) tube, in the range 15-35 eV and for a vanishing momentum transfer
{\bf q} parallel to the tube axis. For this
orientation LFE are negligible.
A strong shift of
the $\pi + \sigma$ plasmon from 22 to 28~eV due to intertube
interactions in the solid can be seen \cite{Gumb}.
The magnitude of this shift reveals a strong dependence of the plasmon
position upon the intertube distance (hence the average valence electron
density) essentially following the plasmon-frequency dependence in the case of
the homogeneous electron gas \cite{Gas}.

This shows that the tubes respond as homogeneous and highly polarizable objects
for parallel {\bf q} in the long-wavelength limit (${\bf q} \to  0$).
A direct consequence would then be that the atomic arrangement, orientation of
bonds and helicity may play a secondary role in the response in this
frequency and {\bf q} range. Therefore, the result for the
$\pi + \sigma$ plasmon (shown in Fig.~\ref{ctubeels}) would be representative
of either of the three tubes since all of them
-- being of nearly the same diameter -- possess the same average electron
density. This is indeed the case as it can be seen in the inset of
Fig.~\ref{ctubeels}
where an almost indistinguishable $\pi + \sigma$ plasmon was also obtained
for the (5,0) tube.
Hence, in contrast to an optical absorption experiment,
small-q loss measurements of {\it the $\pi + \sigma$ plasmon
cannot determine tubular helicities} for a given tube diameter.

The governing factor for the $\pi + \sigma$ plasmon must be traced to the
in-layer graphitic response as it can be seen in Fig.~\ref{ctubeels} where the
loss function for graphite and graphene is also shown
(for in-layer {\bf q} orientation) at comparable (to the tubes) average
electron densities (to within 10 \%) \cite{gra1} (dashed curves in 
Fig.~\ref{ctubeels}).
This shows that the loss function of the tubes for parallel {\bf q} in this
frequency range is
governed by the average-density-dependent part of the in-layer
graphitic response.
Similar plasmon shifts, therefore, can also be expected in other carbon
systems with graphene-based structural blocks e.g. multiwalled tubes.
The present results outline the significance of the {\it $\pi + \sigma$
plasmon as a key measurable spectroscopic quantity} which could
gauge the {\it intertube distances and interactions} in real samples of
carbon nanotubes.

In order to gain a more complete understanding of the in-layer graphitic
response at small q's and how the latter is influenced by the inter-layer
interaction we determined both the loss and dielectric function for
various graphene-like geometries i.e. varying the interlayer spacing or,
equivalently, the (c/a)$_{hex}$ ratio.
In these calculations we also looked at the lower-frequency $\pi$ plasmon.
Bearing in mind the discussion in the
previous paragraph, these calculations could then
serve as benchmarks for predicting the position of the $\pi + \sigma$
plasmon in nanotubes as a function of the intertube distances for low q's.

\begin{figure}[h]
\includegraphics*[width=8cm]{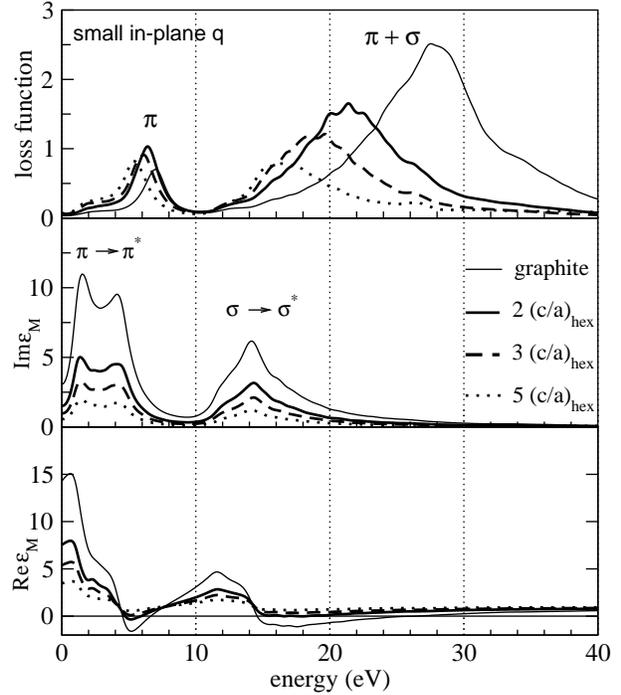}
\caption []
{Loss and dielectric function for small in-plane q
(= 0.22 \AA$^{-1}$) for graphite and
the graphene-like geometries with multiple (c/a)$_{hex}$ ratios.
Only the results without LFE are presented since LFE are negligible
in this q range.}
\label{grapheels}
\end{figure}

The loss and dielectric function for small in-plane {\bf q} (0.22 \AA$^{-1}$)
is shown in Fig.~\ref{grapheels} for graphite and the graphene geometries with
multiple (c/a)$_{hex}$ ratios.
It can be seen that the peak positions of both plasmons
have shifted to lower frequencies when the inter-layer
separation is increased. This effect is very pronounced for the
$\pi + \sigma$ plasmon position.
These results reaffirm that the total ($\pi + \sigma$) plasmon is
extremely sensitive
to the inter-layer interaction for small in-plane q's.
This can be explained as follows: with increasing (c/a)$_{hex}$, i.e.
interlayer spacing, the system becomes an assembly of nearly
isolated graphene sheets. This leads to a decrease of screening
(Re[$\varepsilon_{M}] \to 1$; see Fig.~\ref{grapheels}) with the effect that
$-Im[1/\varepsilon_{M}] \to Im[\varepsilon_{M}]$, namely a coincidence
of the loss and absorption functions.
Since for the in-plane polarization the positions of the absorption peaks do not
change with increasing (c/a)$_{hex}$ (see Im[$\varepsilon_{M}$] in 
Fig.~\ref{grapheels}), then the loss function undergoes important
changes, in particular the $\pi + \sigma$ plasmon. The latter
is displaced at a much faster rate to lower frequencies
towards the 14 eV peak of the absorption spectrum whereas the
$\pi$ plasmon peak is rather insensitive to (c/a)$_{hex}$ since it
is already located very close to the 0-5 eV peak structure of the
$\pi \to \pi^{\star}$ transitions in $Im[\varepsilon_{M}]$.

At present, existing measurements \cite{Kuz,PIC98}
of the loss spectra of samples of single-wall
carbon nanotubes have given a $\pi + \sigma$ plasmon in the frequency
range 21--24 eV for small momentum transfer q.
Our predicted frequency of the $\pi + \sigma$ plasmon for the case of
the distant tubes is within this frequency range (Fig.\ref{ctubeels}).
However, a direct comparison of the present results for the loss function
of (3,3) and (5,0) tubes with the measured loss data
is not straightforward.
The difficulty stems from two factors, tube diameter and alignment, which
have a competing effect on the $\pi + \sigma$ plasmon position:
a) Bulk samples of single-wall tube material possess a mean diameter of
14 \AA~\cite{PIC98}, which is considerably larger to the range of 4 \AA~
studied in the present work.
Assuming a common intertube spacing, then larger diameters would
give rise to a displacement of the $\pi + \sigma$ plasmon
towards lower frequencies since the diametrically-opposed wall parts of
the same
tube are facing each other at larger distances, i.e. a situation
resembling
a stacking of graphene sheets
with larger (c/a)$_{hex}$ ratios (see Fig.~\ref{grapheels}). For instance,
the 5 (c/a)$_{hex}$ ratio corresponds to an intersheet separation of 16.8
\AA~ and the corresponding $\pi + \sigma$ plasmon peak is at 16 eV.
b) The alignment of the tubes in the samples is not perfect; therefore
it is to be expected that also out-of-plane excitations of graphitic
origin
will contribute to the response. These excitations \cite{gra1}
should tend to produce
a more diffuse shape for the loss spectrum,
heavily dampening the $\pi + \sigma$ plasmon and shifting
the observed peak towards higher frequencies.

\begin{figure}[h]
\includegraphics*[width=8cm]{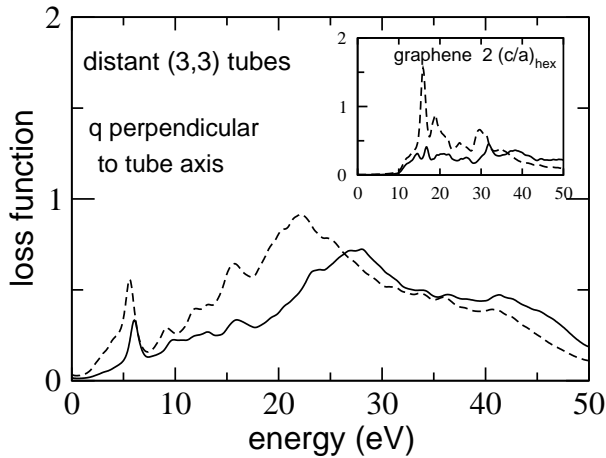}
\caption []
{Calculated RPA loss function for the carbon (3,3) distant tubes for
small {\bf q} (0.19 \AA$^{-1}$) of orientation perpendicular to the tube axis.
In the inset the loss function of graphene \cite{gra1} for 2 (c/a)$_{hex}$ and
for a similar {\bf q} of orientation perpendicular to the sheet is displayed.
Continuous and dashed curves denote results obtained with and without LFE,
respectively.}
\label{ctubeels2}
\end{figure}

The type of response described just above can be clearly seen in 
Fig.~\ref{ctubeels2} which shows the loss function of the distant (3,3) 
tubes for {\bf q} orientation perpendicular to the tube axis.
For this orientation in-plane as well as out-of-plane graphene
excitations contribute to the tube response. The latter cause the
diffuse shape of the loss function (see inset of Fig.~\ref{ctubeels2}). 
LFE are now very important and the
peak position of the $\pi + \sigma$ plasmon is at 28 eV.

\begin{figure}[h]
\includegraphics*[width=8cm]{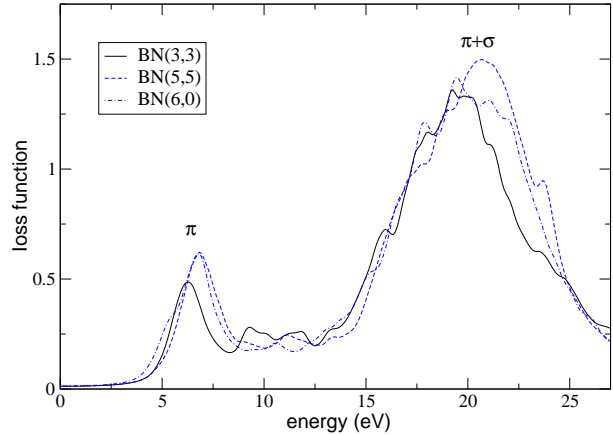}
\caption []
{Loss spectra for three different BN nanotubes with momentum
transfer $q \rightarrow 0$ along the direction of the tube axis.
The spectra are calculated with an inter-tube distance of 7.4 {\AA}.
Since the spectra with and without LFE are almost indistinguishable,
only the calculation without LFE is shown.}
\label{fig9}
\end{figure}

For completeness, we show in Fig.~\ref{fig9} also the calculated
EELS spectra for small BN nanotubes with momentum transfer $q \rightarrow 0$
along the tube axis. As in the case of carbon tubes, two main features
are clearly pronounced: the $\pi$-plasmon at 6-7 eV and the high energy
collective oscillation ($\pi+\sigma$ plasmon) around 20 eV.
The exact position of the peaks depends on the radius and chirality
of the tubes and on the intertube distance.
In order to compare with experimental EELS-spectra on multi-wall
BN tubes \cite{fuentes}, an extrapolation to larger-diameter tubes is needed
and the inter-wall interactions have to be taken into account.

\section{Discussion and Conclusions}
\label{finsec}

In this closing section we should like first to comment on the validity
of RPA and TDLDA for the description of the optical spectra of the tubes.
It is well known that RPA and TDLDA often give excellent results for
{\it loss} spectra \cite{REV}. However, in this work we have seen
that also very good {\it absorption} spectra were obtained for the carbon tubes,
even at the RPA level, despite the fact that both RPA and TDLDA are known
to fail badly in the description of
absorption spectra of many bulk materials (silicon and argon being two
representative cases) \cite{REV}.
Our explanation in this regard is the following:
for the light polarization parallel to the tube axis, the ensuing
screening is significant and, therefore, XC effects are damped.
For the perpendicular polarization and for
the larger inter-wall distance ($D_{t}=$ 5.5 \AA) the
tubes behave essentially like isolated systems, where strong
cancellations are known to occur between
self-energy corrections and the electron-hole interaction, i.e.
between XC effects (see Ref.\cite{REV}).
The experimental precision is then not high enough
(also due to the almost vanished absorption intensity in the relevant
frequency range)
to discern to which extent XC effects should be better described
by approximations beyond the TDLDA.
In view of these considerations, we can be confident regarding the quality
of the calculations.
On the other hand, the results for the strongly
interacting tubes (smaller intertube distance) should be regarded as
qualitative since the system becomes then more similar to a
solid where the cancellations may be more incomplete, and since no
direct comparison to experiment is possible, at present, in this case.

Still, excitonic effects are not included in the calculations.
They could play a role in both carbon and BN nanotubes,
leading to redistribution of oscillator strength and/or appearence of new
peaks in the bandgap (bound excitons). They might be the reason for the
anomalous $E_{11}/E_{22}$ ratio~\cite{mele}
measured recently for carbon tubes~\cite{bac03}.
We expect a stronger deviation of measured optical spectra from
theoretical ones in the case of BN nanotubes. For this large band
gap material, two effects will most likely play an important role:
Quasi-particle corrections will widen the band gap as in the case
of hexagonal BN, where a GW calculation has demonstrated a band-gap
increase from 4 eV to 5.5 eV \cite{bnblase}. Excitonic effects, in contrast,
will lead to isolated states in the band gap or to an overall reduction
of the band-gap. To which extent quasi-particle corrections and excitonic
effects cancel each other is presently not clear. Work along these
lines is in progress~\cite{ludger}.
For the role of quasi-particle corrections and
excitonic effects in carbon tubes, we refer the reader to Ref. \cite{excilou}.

In conclusion, the present {\it ab initio}
calculations of the optical absorption of
small-diameter carbon and BN nanotubes give good agreement with the
available experimental data.
The inclusion of local field effects in the response to a perturbation
with perpendicular polarization is necessary for
a proper description of the depolarization effect leading to a
suppression of the low-energy absorption peaks for both types of tubes.
This suppression can also explain recent findings in near-field Raman 
microscopy of
single-wall carbon nanotubes~\cite{achim}. The proper analysis of the
polarization dependence of the absorption cross section is very important
in order to describe the surface enhanced Raman scattering experiments
in isolated carbon nanotubes~\cite{jorio,duesberg}.
In carbon tubes the position of the first absorption peak strongly varies
with the tube indices while in BN tubes the first peak is determined
by the band gap of BN and is therefore mostly independent of $(n,m)$.
For the BN tubes some of the fine-structure which distinguishes
tubes of different chirality is only visible in the UV
region which gives rise to the hope that this energy regime will
be probed in the future.

The intertube interaction was also found to be very important.
For the carbon
tubes this interaction is the governing factor which determines the
position of the higher-frequency $\pi + \sigma$ plasmon.
This plasmon, hence, may prove to be a very useful spectroscopic
quantity probing intertube interactions and distances in real samples.
Finally, the corresponding results for the layered constituents
--- graphene and BN sheets --- revealed that some aspects of the
tubular dielectric response can be explained even
at a quantitative level from the in-layer
and interlayer response of the sheets.

\section{Acknowledgments}
This work was supported by the European Community Research Training
Networks NANOPHASE (HPRN-CT-2000-00167) and COMELCAN (HPRN-CT-2000-00128),
by Spanish MCyT(MAT2001-0946) and University of the Basque Country
(9/UPV 00206.215-13639/2001).
The computer time was granted by IDRIS (Project No. 544), DIPC 
and CEPBA (Barcelona).
The authors also gratefully acknowledge fruitful discussions with Thomas
Pichler and Nathalie Vast.


\begin{thebibliography}{99}
\bibitem{optical} Combined scanning-tunneling microscopy and spectroscopy 
(STM/STS) is a very
reliable tool for local tube characterization~\cite{STS}, however it cannot be
used for routine tube mapping.
\bibitem{nvalues}
Typically, for the systems and excitation energies treated in this work, it 
was sufficient to compute bands up to $N=4n_{occ.}$, where $n_{occ.}$ is the 
number of occupied bands. In certain cases, especially for evaluating the
loss function, we included more unoccupied bands in the summations.
\end{thebibliography}
\end {document}